\documentclass[prb,preprint]{revtex4-1}

\usepackage{amsmath}  
\usepackage{amsfonts} 
\usepackage{graphicx} 
\usepackage[usenames, dvipsnames]{color}
\usepackage{eurosym}
\usepackage{gensymb}

\begin{document}


\title{Temperature-dependent transport measurements with Arduino}

%
%
%


\author{A. Hilberer}
\author{G. Laurent}
\author{A. Lorin}
\author{A. Partier}
\affiliation{Magist\`{e}re de Physique Fondamentale, D\'{e}partement de Physique, Univ. Paris-Sud, Universit\'{e} Paris-Saclay, 91405 Orsay Campus, France}

\author{J. Bobroff}
\author{F. Bouquet}
\author{C. Even}
\affiliation{Laboratoire de Physique des Solides, CNRS, Univ. Paris-Sud, Universit\'{e} Paris-Saclay, 91405 Orsay Campus, France}

\author{J. M. Fischbach}
\affiliation{Magist\`{e}re de Physique Fondamentale d'Orsay, D\'{e}partement de Physique, Univ. Paris-Sud, Universit\'{e} Paris-Saclay, 91405 Orsay Campus, France}

\author{C. A. Marrache-Kikuchi}
\email{claire.marrache@u-psud.fr}
\affiliation{CSNSM, Univ. Paris-Sud, CNRS/IN2P3, Universit\'{e} Paris-Saclay, 91405 Orsay, France}

\author{M. Monteverde}
\affiliation{Laboratoire de Physique du Solide, CNRS, Univ. Paris-Sud, Universit\'{e} Paris-Saclay, 91405 Orsay Campus, France}

\author{B. Pilette}
\affiliation{Magist\`{e}re de Physique Fondamentale, D\'{e}partement de Physique, Université Paris-Sud, Universit\'{e} Paris-Saclay, 91405 Orsay Campus, France}

\author{Q. Quay}
\affiliation{Laboratoire de Physique des Solides, CNRS, Université Paris-Sud, Universit\'{e} Paris-Saclay, 91405 Orsay Campus, France}

\begin{abstract}
The current performances of single-board microcontrollers render them attractive not only for basic applications but also for more elaborate projects, amongst which are physics teaching or research. In this article, we show how temperature-dependent transport measurements can be performed using an Arduino board, from cryogenic temperatures up to room temperature or above. We focus on two of the main issues for this type of experiments: the determination of the sample temperature and the measurement of its resistance. We also detail two student-led experiments: evidencing the magnetocaloric effect in Gadolinium and measuring the resistive transition of a high critical temperature superconductor.
\end{abstract}

\maketitle 

\section{Introduction}
The development of single-board microcontrollers and single-board computers has given physicists access to a large variety of inexpensive experimentation that can be used either to design simple test benches, put together set-ups for class demonstration or devise student practical work. Moreover, specifications of single-board components are now such that, although they cannot rival with state-of-the-art scientific equipment, one can nonetheless derive valuable physical results from them.

We will here focus on the Arduino microcontroller board\cite{Arduino}. Let us note that other boards, such as MBED, Hawkboard, Rasberry Pi, or Odroid to cite but a few, exist which may be cheaper and/or have better characteristics than Arduino. In our case, we have employed Arduino boards to take advantage of the important users community. This has been an important selling point for the students with whom we are working.

Indeed, our experience with Arduino is primarily based on undergraduate project-based physics labs\cite{Bouquet2016} we have initiated within the Fundamental Physics Department of Universit\'{e} Paris Sud for students to gain a first hands-on practice of experimental physics. In these practicals, students are asked to choose a subject they want to study during a week-long project. They then have to design and build the experiment with the equipment available in the lab. The aim is not only to study a physical phenomenon, but to do so using inexpensive materials and low-cost boards.

In this article, we will describe two projects that have been developed by third year students. The first one aimed at quantifying the magnetocaloric effect in Gadolinium and the second one, which has been popular amongst students, consisted in measuring the resistive transition of a high critical temperature superconductor (HTCS). However, the techniques to do so can more generally be used for any experiment involving the measurement of a low voltage while varying the set-up temperature. In particular, they could be applied to simple transport characterization of samples in research laboratories.

In the following, we will focus on two important issues for this kind of measurements: thermometry and thermal anchoring on the one hand and measuring resistances on the other.

\section{Determining the temperature of an object using microcontrollers}
One of the experimental control parameters that is most commonly used to make a physical system properties vary is temperature. There is a wide variety of temperature sensors, depending on the temperature range of interest. The aim of this paper is not to list those, but rather to focus on the most ordinarily found sensors compatible with an Arduino read out. We will also review some basic techniques to ensure a proper thermal contact between the sample and the thermometer.

\subsection{Sensors types}

\subsubsection{Built-in Arduino sensors}

There are a number of temperature sensors that are generally sold with standard Arduino kits. The Arduino Starter Kit, for instance, comes with a TMP36 low voltage temperature sensor\cite{TMP36} (available for about \$1 if purchased separately) which operating principle is based on the temperature-dependence of the voltage drop across a diode.

The advantage of this type of thermometer is that it can be directly plugged into Arduino without any additional electrical circuit. Furthermore, provided that the corresponding library is downloaded, the temperature is straightforwardly read via the computer interface in \degree C, so that no calibration is needed.

However, these sensors are limited in accuracy and operation: the TMP36 sensor for example has a $\pm 2$\degree C precision over the $-40$\degree C to $+125$\degree C range where it can operate. If they are extremely convenient for non-demanding temperature read-outs, such as students atmospheric probes for example\cite{Station_meteo}, they are not adapted to the precision needed for most research lab experiments.


\subsubsection{Thermocouple}

Thermocouples are cheap and robust thermal sensors that are industrially available for about \$15, and which cover a wide range of temperatures (for example from $-200$\degree C to $+1250$\degree C for a type K thermocouple\cite{Thermocouple}). They are one of the few thermometers that are reliable at temperatures much higher than room temperature.

Thermocouples are also extremely convenient to measure the temperature of small-sized samples. Indeed, only the hot junction between the two metals needs to be in contact with the region where the temperature is to be monitored. On the other hand, the quality of the readings will strongly depend on how thermally stable the cold junction is, and the temperature measurement is less precise than with a thermistor. Indeed, the voltage to be measured is small: the sensitivity of a thermocouple is of the order of tens of microvolts per Kelvin, and it decreases when the temperature decreases (a type-K thermocouple has a sensitivity of 40~$\mu$V.K$^{-1}$ at room temperature but of 10~$\mu$V.K$^{-1}$ at liquid nitrogen temperature).

Let us note that an amplification of the voltage signal is then needed to read the temperature with Arduino. Some chips provide a ready-to-use thermocouple amplifier for microcontrollers (such as the MAX31856 breakout with a resolution of a quarter of a kelvin when using the Adafruit library \cite{Thermocouple_ampli} and an accuracy of a few kelvins). Better sensitivity could be achieved with a home-made amplifier (see below) and some care.


\subsubsection{Platinum thin resistive films}
Platinum thin films are practical and very reliable resistive thermometers typically working from 20 K to 700 K\cite{Pt100_spec}. They are therefore suited for cryogenic applications -- at least down to liquid nitrogen temperatures -- as well as for moderate heating. The advantage of this sensor is that its response is entirely determined by the value of its resistance at $0$\degree C\cite{Pt100}. The most commonly used platinum resistance is the so-called Pt100 which has a resistance of 100 $\Omega$ at $0$\degree C and costs approximately \$3 to \$5. These thermal sensors have a typical precision of about 20 mK up to 300 K and about 200 mK above room temperature. Moreover, their magnetic field-dependent temperature errors are well-known\cite{Pt100_spec}.

It is possible to mount those resistances on a dedicated Arduino resistance-to-temperature converter such as MAX31865\cite{MAW31865}, but it is often simpler to plainly measure the resistance with a dedicated electrical circuit as will be explained in section \ref{Sec:Small_V} This is particularly convenient for low or high temperature measurements for which the Arduino board cannot be at the same temperature as the thermometer and the sample.

\subsection{Thermal anchoring}

For the temperature measurement to be relevant, the thermometer must be in good thermal contact with the sample. How to achieve a good thermal anchoring is a subject of investigation in itself, but in this section we will outline a few standard techniques, focusing on the low temperature case.

To cool down a sample at low temperatures, one could use a Peltier module, but the simplest -- and not so expensive -- way is to use liquid nitrogen. Some basic safety measures have to be taken to manipulate this cryogenic fluid: use protection glasses, gloves, work in a well-ventilated room and, above all, ensure that it is poured into a vessel that is not leak-tight to allow natural evaporation of liquid nitrogen and avoid pressure build-up in the vessel. Once these precautions are observed, the manipulation itself is relatively safe.

To ensure that the thermometer indeed probes the sample temperature, the most obvious technique is to solidly attach the thermometer to the sample using good thermal conductors. The thermal sensor can for instance be glued onto a copper sample holder. The glue then has to retain its properties at the probed temperature range. In the low temperature case, one frequently uses GE 7031 varnish which sustains very low temperatures and can easily be removed with a solvent. Alternatively, the thermometer could be mechanically fixed with a spring-shaped material which elasticity is maintained at low temperature, such as CuNi sheets. Upon cooling, the spring-shaped material will continue to apply pressure onto the thermometer, thus ensuring a good mechanical and thermal contact with the sample holder.

Another method is to thermally insulate the thermometer and the sample from the outside world, while putting them in contact with a common thermal bath. This can be done by inserting them into a container filled with glass beads of a few millimeters in diameter\cite{Ireson} (inset of figure \ref{fig_4_point_ampli_supra}), or, alternatively, sand. These materials provide a good thermal insulation of the \{sample+thermometer\} system from the outside world while allowing for an important thermal inertia. Moreover, when working at temperatures down to 77 K, they limit the liquid nitrogen evaporation so that the temperature increases back to room temperature only very slowly: typically for a volume 1 L of beads that is initially immersed in liquid nitrogen, the temperature reaches back 300 K in 3 or 4 hours. The heat exchange between the sensor and the sample is then guaranteed through the evaporated N$_2$ gas, thus ensuring the temperature is homogeneous within the entire volume. An alternative method for achieving good thermal contact between the sensor and the sample through gas exchange is explained in reference \cite{Rossano}.


\section{Measuring resistances with microcontrollers}
\label{Sec:Small_V}
Microcontroller inputs give a reading of electric potentials. Measuring resistances is then slightly more complicated than plugging a resistance into an ohmmeter. For educational purposes, this is actually rather valuable since it gives students the opportunity to experiment with the notion of resistance and to realize that even the simplest measurement may present some challenge. In the following, we will present standard methods to measure resistances and will particularly focus on the low-resistance case.

\subsection{Current-Voltage measurement}
\begin{figure}
\begin{center}
\includegraphics[width=0.4\textwidth]{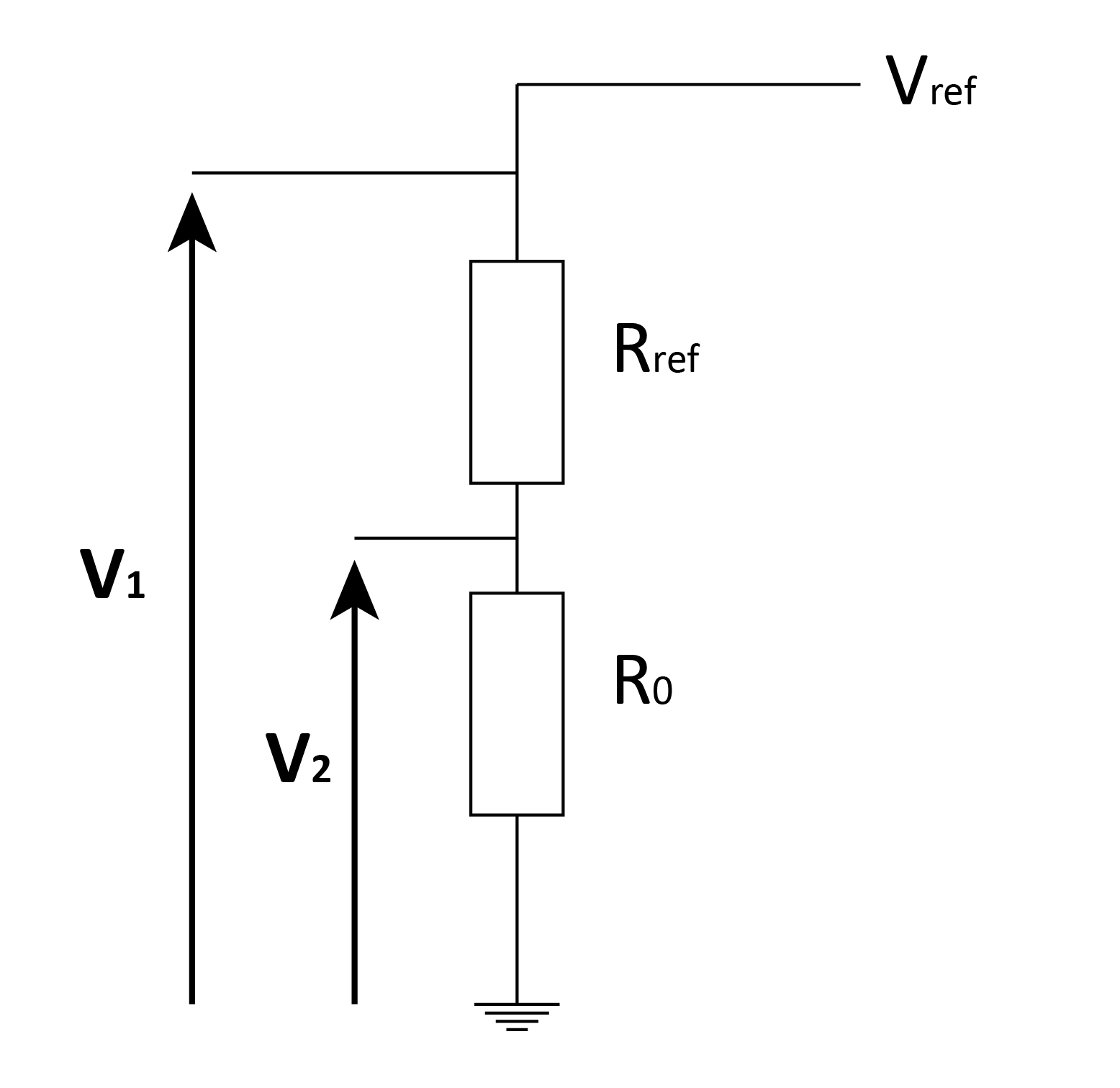}
\end{center}
\caption{Schematic representation of the current-voltage set-up.} \label{fig:Diviseur_tension}
\end{figure}

The simplest set-up for measuring a standard resistance is the voltage divider set-up represented in figure \ref{fig:Diviseur_tension}: the resistance of interest $R_0$ is put in series with a reference resistance $R_{ref}$. The voltage drop across both resistances is controlled by the board 5 V output. The potential $V_1$ can be read by one of the microcontroller's inputs and should be close to 5 V. The potential $V_2$ -- read by a second input -- corresponds to the voltage drop across the unknown resistance. $R_0$ can then be determined through the simple relation:

\begin{equation}
R_0=\frac{V_2}{V_1-V_2}R_{ref}
\end{equation}
\noindent The monitoring of $V_1$ allows for a better precision through a direct monitoring of the current. Arduino's 5 V output sometimes varies in time. To have a better stabilization of the voltage, it may be useful to use an external power source for the microcontroller and not use the computer's USB output. Let us note that, if $R_{ref}\gg R_0$, the current through the circuit can be considered to be constant, which is often very convenient when the resistance measurement does not require a large precision.

This method presents several drawbacks when dealing with small values of $R_0$: since the ultimate resolution of an Arduino UNO board is of about 1 mV with $V_{ref}=1.1$ V, one cannot measure $R_0$ smaller than about $2\times10^{-4}R_{ref}$. In the case of a standard commercial HTCS sample for instance, the normal state resistance is often of the order of a few tens of m$\Omega$. To observe the resistance drop across the critical temperature $T_c$ of a superconductor, $R_{ref}$ should then be of the order of a few $\Omega$. Such resistances are commercially available or, alternatively, can be custom-made with a relatively good precision (of the order of a few m$\Omega$) by using a long string of copper wire (commercially available Cu wires of 0.2 mm in diameter have a resistance of about 0.5 $\Omega$/m for example). However, unless $V_2$ is amplified, the precision of the measurement is not optimal. Moreover, using this method to measure small resistances leads the circuit current to exceed the maximum current allowed at the microcontroller's output. In the following, we will see another method to measure small resistances.

\subsection{Wheatstone bridge}
\begin{figure}
\begin{center}
\includegraphics[width=0.4\textwidth]{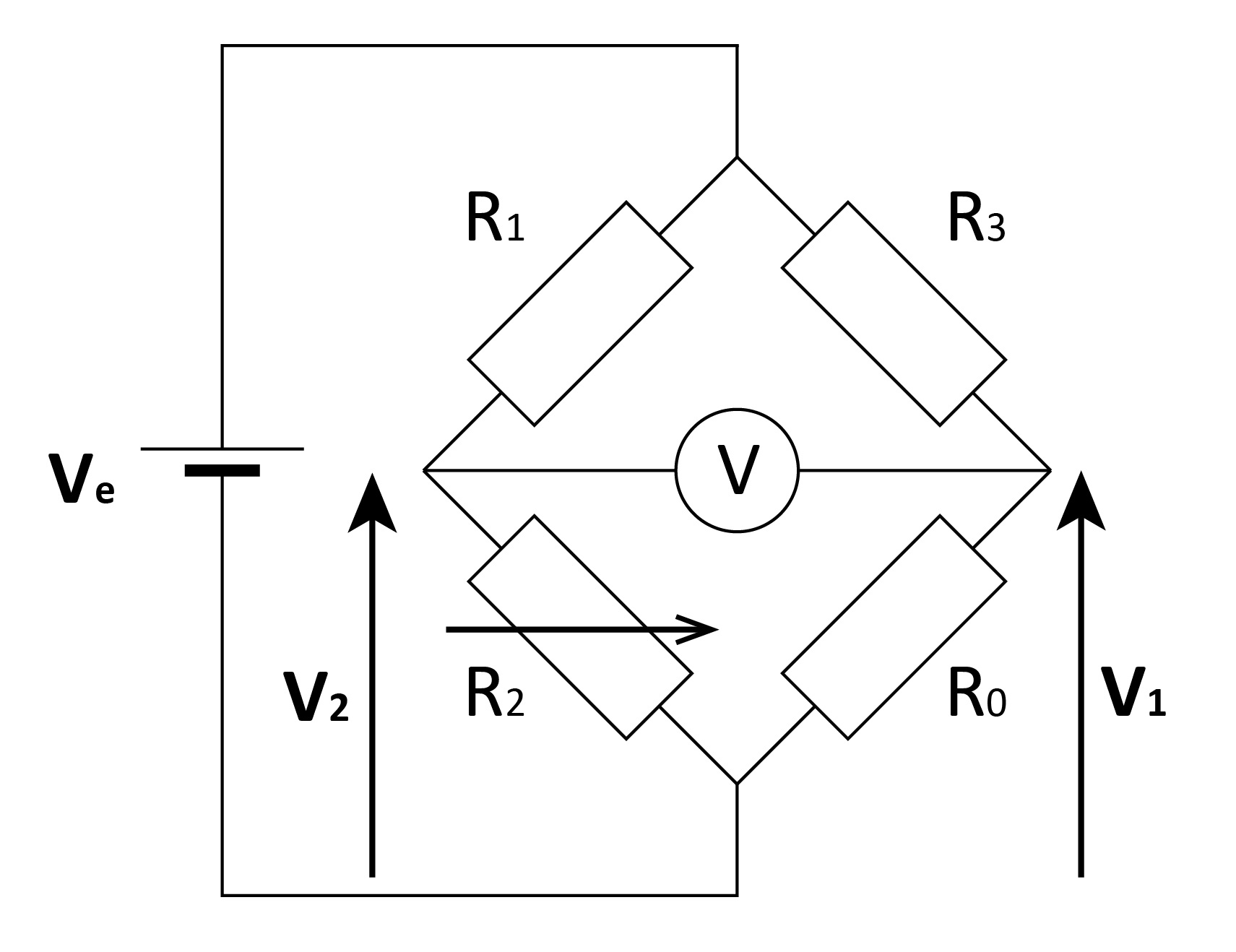}
\end{center}
\caption{Schematic representation of the Wheatstone bridge: $R_1$ and $R_3$ are fixed resistances, $R_2$ is a tunable resistance, and $R_0$ is the resistance of interest.} \label{fig_Wheatstone}
\end{figure}

Another resistance determination method which can achieve a good precision is the Wheatstone bridge. The principle of the measurement is illustrated in figure \ref{fig_Wheatstone}. $R_1$ and $R_3$ are fixed value resistances, while $R_2$ is a tunable resistance and $R_0$ the resistance of interest. The potentials $V_1$ and $V_2$ are then related by:
\begin{equation}
V_2-V_1 = \left(\frac{R_2}{R_1+R_2}-\frac{R_0}{R_0+R_3}\right)V_e
\label{eq:Wheatstone}
\end{equation}
\noindent The bridge is co-called ``balanced'' when $R_2$ is tuned such that $V_1$ and $V_2$ are equal. The resistances are then related through:
\begin{equation}
R_0=\frac{R_2R_3}{R_1}
\end{equation}
\noindent The precision that can be achieved through this method and using a microcontroller is about the same as for the current-voltage measurement method. However, this method is not very practical when dealing with resistances $R_0$ that vary, since the bridge has to be maintained close to balance at each measurement point. In particular, it is not well suited for the measurement of a superconductor's resistive transition.

\subsection{Voltage amplifier}
\begin{figure}
\begin{center}
\includegraphics[width=0.4\textwidth]{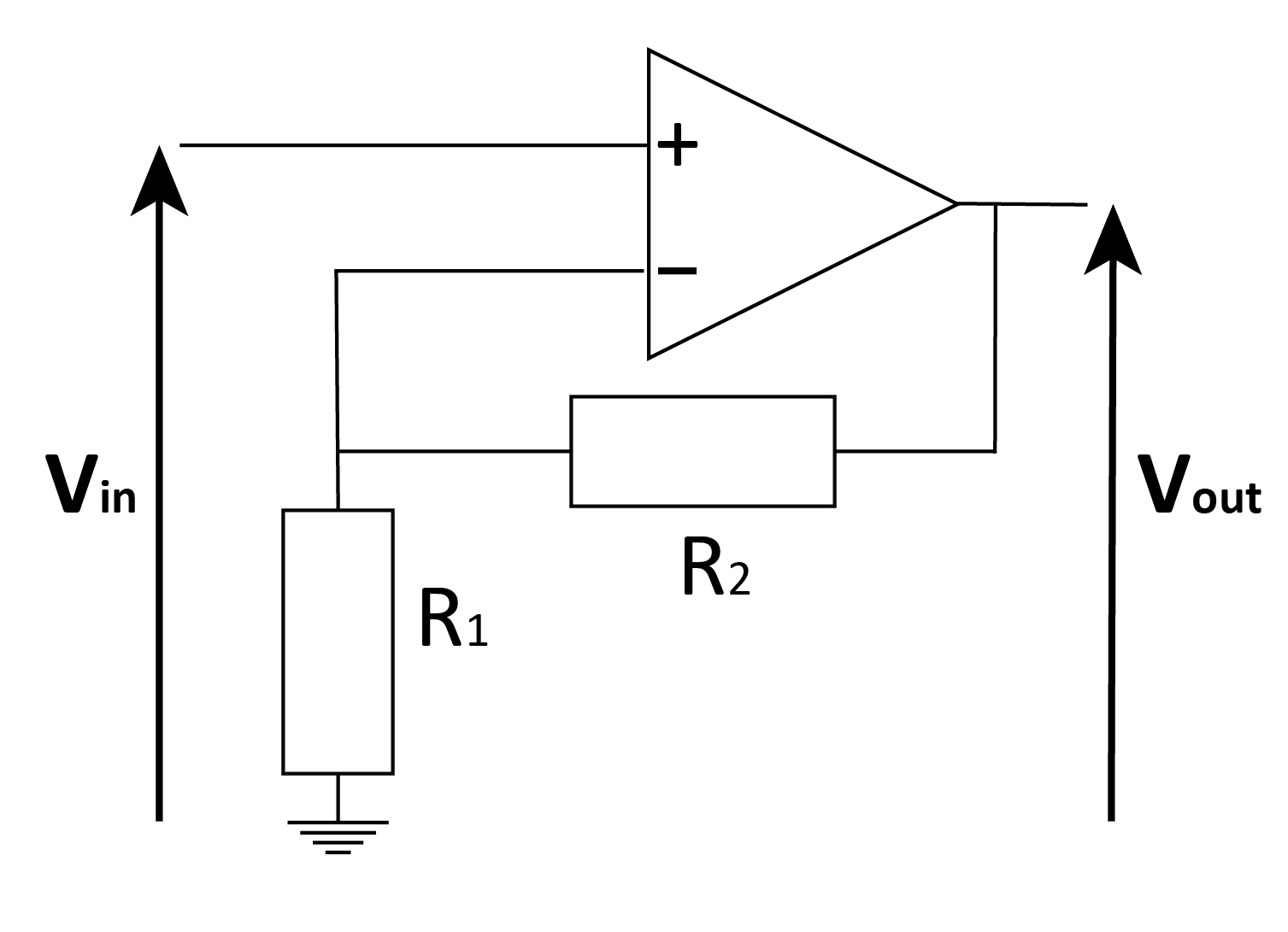}
\end{center}
\caption{Voltage amplification.} \label{fig_AO.jpg}
\end{figure}

The most practical solution for the measurement of small voltages -- and hence small resistances -- is the amplification of the potential difference across the resistance. This can be done via standard voltage amplification set-ups using operational amplifiers either in single-ended or differential input configurations. In the single-ended case, illustrated in figure \ref{fig_AO.jpg}, the output potential is given by:
\begin{equation}
V_{out}=1+\frac{R_2}{R_1}V_{in}
\end{equation}
\noindent The input voltage $V_{in}$ can then be amplified at will, depending on the ratio $\frac{R_2}{R_1}$. The output voltage $V_{out}$ can then be read by the microcontroller.

This amplification method has a much larger precision than the previously mentioned methods, does not require tuning at each data point and can be used to measure any small voltage: the voltage drop across a superconductor, but also the difference of potential across a thermocouple, or to derive thermoelectric coefficients (Seebeck or thermopower).

Going beyond this simple amplification method requires substantially more work. One possible method is to fabricate a microcontroller-based lock-in amplifier, as demonstrated in reference \cite{Detec_synch}.

\subsection{Using another DAC than Arduino's}
The specifications of Arduino Digital-Analog Converter are often the main limitation in the above measurements. As already mentionned, the Arduino ADC provides -- at best -- 10 bits on the 1.1 V internal reference voltage, and can only measure a voltage in single-ended configurations.

One alternative would be to use another microcontroller, with a better ADC. For example, the low-cost FRDM-KL25Z from NXP~\cite{refNXP} provides a ADC that can measure a voltage either in single-ended or in differential input configurations with 16 bits on 3.3 V.

The ease-of-use and the large users community can be a strong motivation to keep Arduino as your board of choice. In which case, a second solution would be to use an external ADC when better resolution or a differential configuration is needed. For example, we have tested the ADS1115 chip~\cite{refAda}: this external ADC can measure 4 single channels or 2 differential channels with 16 bits on 4.1 V. The possibility of a preamplification up to 16 times brings the resolution down to 8 $\mu$V per bit instead of the standard 5 mV (or 1 mV with the 1.1 V internal reference).

The possibility of measuring a voltage in a differential configuration with a resolution better than 10 $\mu$V are two important advantages that open many interesting possibilities for physics measurements: for instance, measuring a strain gauge or a resistance in a four-wire configuration, or measuring directly a thermocouple or the resistance of a superconductor across the transition.

The main drawback of this method is that it is not as easy as using the Arduino ADC: a library should be installed first (but good tutorials can be found online, see for example Ref.~\cite{refAda}). Also, an external ADC is generally not as robust as Arduino's and the user should carefully monitor the voltage input so as not to damage the ADC.



\section{Evidencing a magnetocaloric effect with microcontrollers}
\label{Sec:magnetocaloric}
To illustrate these methods, let us detail the magnetocaloric effect that two students\cite{AH-GL} have measured. This effect consists in the temperature change occurring when a magnetic material is placed in a varying magnetic field. A more detailed explanation of the phenomenon can be found in reference \cite{Magnetocaloric}.

In our case, Gadolinium (Gd) was chosen for its paramagnetic properties and its Curie temperature close to room temperature ($T_{Curie}=292$ K). At a temperature of about 298 K, a 2.242 g Gd sample was submitted to the magnetic field created by a neodymium magnet of maximum value 0.51 T.

\begin{figure}
\begin{center}
\includegraphics[width=0.52\textwidth]{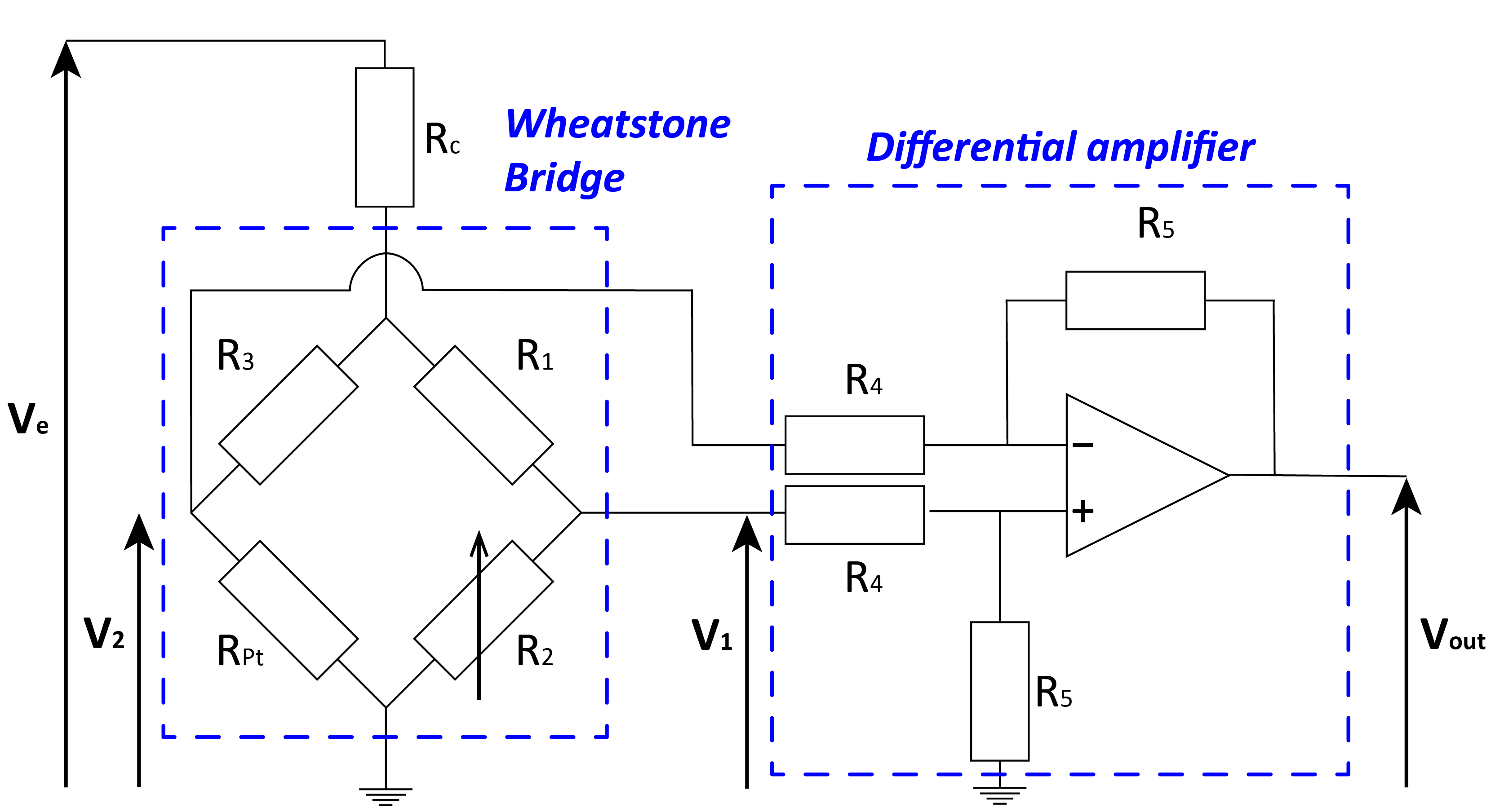}
\end{center}
\caption{Measurement of the resistance of a Pt resistive thermometer with a Wheatstone bridge and amplified by an opamp-based circuit.} \label{fig:Ampli_magneto}
\end{figure}

In this experiment, the challenge was to measure the small temperature difference induced by the application of a magnetic field. To this effect, a Pt100 thermistor was put in good thermal contact with the Gd sample via thermal paste. Its resistance change was measured by a Wheatstone bridge with the following characteristics: $R_1=R_3=100$ $\Omega$, $R_2$ has been set at 108 $\Omega$ to be close to balance at the considered temperature and $V_e=5$ V via Arduino's internal source. An additional resistance $R_c=800~\Omega$ was placed in series to limit the current going through the Pt100, thus avoiding heating the thermometer. $V_e$ is then replaced by $\frac{R_1+R_2}{R_1+R_2+2R_c}V_e$ in equation \ref{eq:Wheatstone}. The off-balance difference of potential $V_2-V_1$ was differentially amplified with a gain of 100 ($R_4=1.5$ k$\Omega$ and $R_5=150$ k$\Omega$). The voltage $V_{out}$ was then read by the board (Arduino Mega in this case) using 2.56 V as Arduino's ADC reference voltage \cite{Arduino_int_ref}. The overall read-out circuit is schematically shown in figure \ref{fig:Ampli_magneto}. The temperature is then inferred knowing that, in the [273 K - 323 K] range, the Pt100 response can be linearly fitted by:
\begin{equation}
 T[ \text{K}] = 2.578 R_{Pt}[\Omega]+15.35
\end{equation}

\begin{figure}
\begin{center}
\includegraphics[width=0.5\textwidth]{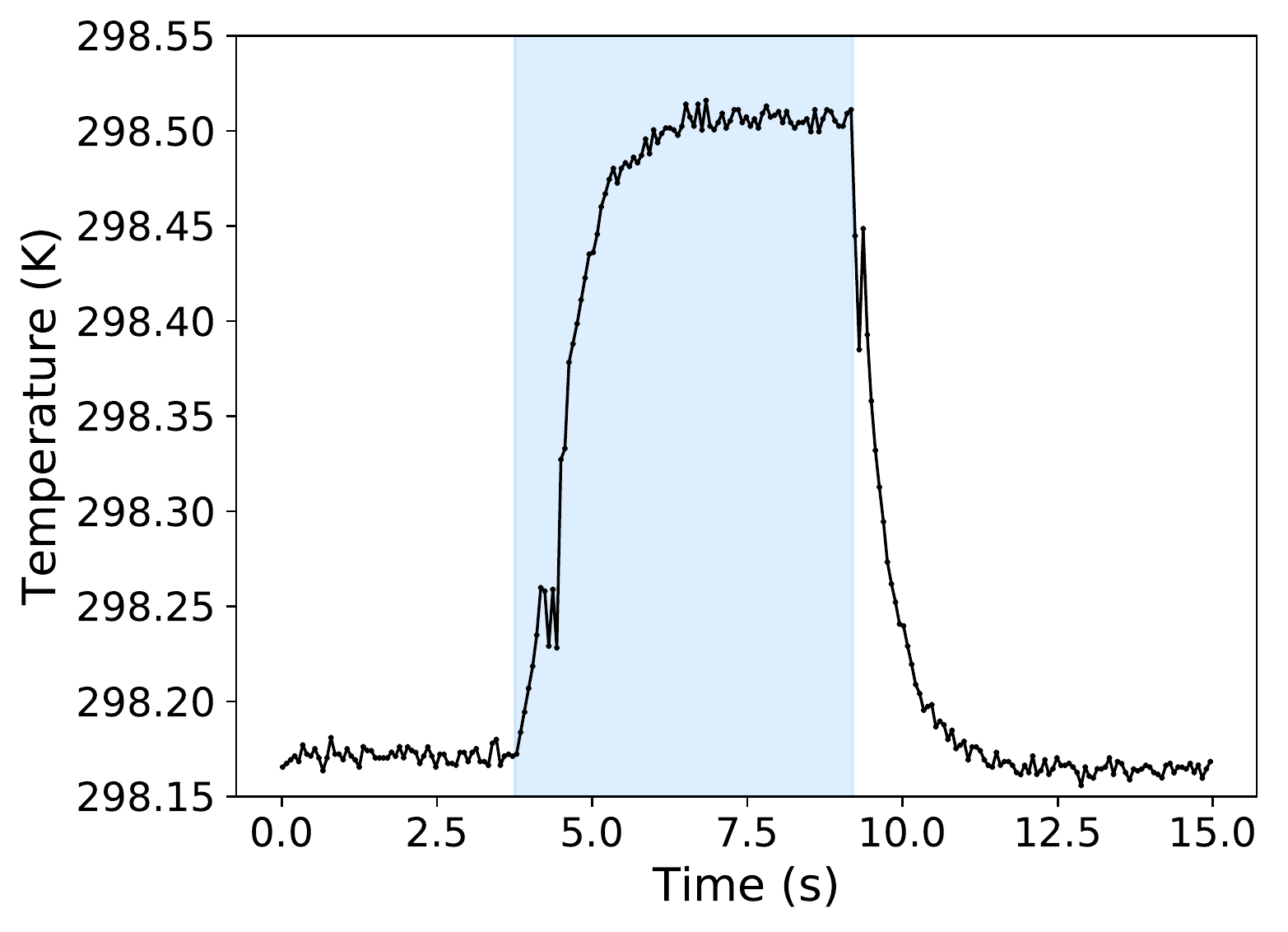}
\end{center}
\caption{Magnetocaloric effect in a Gd sample submitted to a 0.51 T magnetic field (blue background) before going back to the zero-field situation (white background). Each data point corresponds to the average of 50 measurements. The noise level is of the order of 10 mK.} \label{fig:MC_effect}
\end{figure}

As illustrated in figure \ref{fig:MC_effect}, the magnetocaloric effect is clearly visible with an amplitude of about $\Delta T\simeq 0.33\pm0.01$ K and a time scale of a few seconds. The resolution of the setup corresponds to 50 mK (18 m$\Omega$). Each data point of figure \ref{fig:MC_effect} corresponds to an average of 50 measurements so that the effective noise that can be observed is of about 10 mK, or about 5 m$\Omega$ in resistance. This yields a relative precision for the measurement of a few $10^{-5}$ which is remarkable given the simplicity of the apparatus . When the magnet is taken away from the Gd sample, the temperature decreases back to its initial value, as predicted by the isentropic character of the magnetocaloric effect.

\section{Measuring a superconducting resistive transition with microcontrollers}
\label{Sec:Supra}
The second experiment we would like to detail is the measurement of the superconducting resistive transition of a HCTS. Indeed, in such compounds, the critical temperature $T_c$ below which the sample is superconducting and exhibits zero resistance is larger than 77 K. The transition can therefore easily be observed by cooling the sample down to liquid nitrogen temperature and warming it back up to room temperature.

\begin{figure}
\begin{center}
\includegraphics[width=0.5\textwidth]{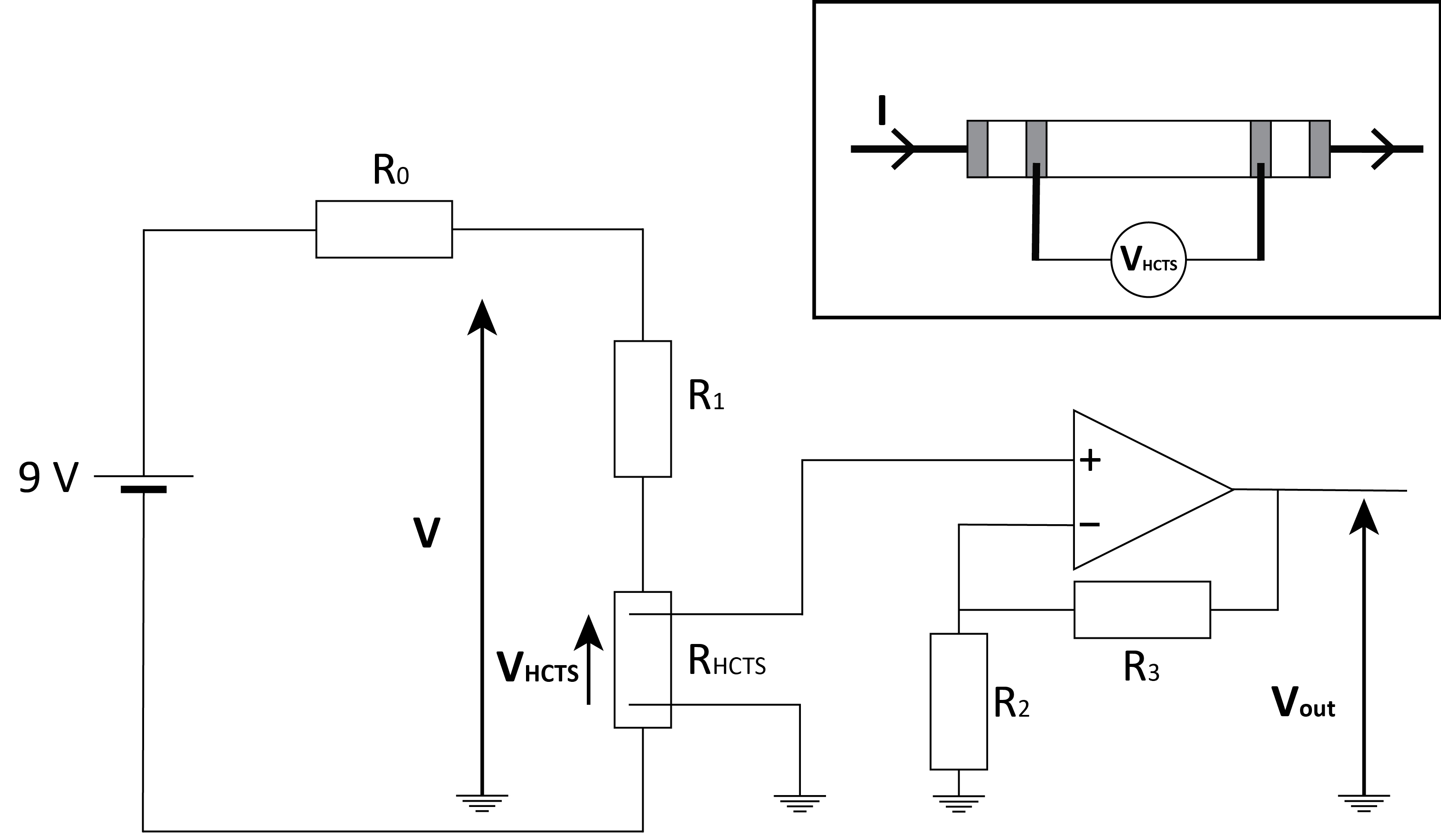}
\end{center}
\caption{Amplification of the voltage drop across a superconducting sample. Inset: geometry of the HCTS sample.} \label{fig:supra_montage}
\end{figure}

In the present case, the HCTS is a commercial Bi$_2$Sr$_2$Ca$_2$Cu$_3$O$_{10}$ sample which specifications indicate a critical temperature $T_c=110$ K at mid-transition point and a room temperature resistivity of $1$ m$\Omega$.cm\cite{supra-ref}. In this case, the experimental challenge is therefore to measure very small resistances with a good precision. To achieve this, it is essential to adopt a four-wire measurement configuration for the superconductor, as schematized in the inset of figure \ref{fig:supra_montage}. Indeed, in this way, no contact resistances or connection wires contribute to the measured resistance. Moreover, the voltage drop across the superconductor has been amplified by a factor of 480 by a single-ended operational amplifier set-up as shown in figure \ref{fig:supra_montage}. The current going through the superconductor is fixed by $R_0=110.0$ $\Omega\gg R_{1},R_{HCTS}$  and is experimentally measured via the potential $V$ read at the extremity of a home-made resistance $R_{1}=1.18$ $\Omega$, made out of copper wire.

The temperature has been measured with a Pt100 resistive thermometer using the set-up shown in figure \ref{fig:Diviseur_tension} with $R_{ref}=217.3$ $\Omega$ and a reference voltage of $V_{ref}=3.3$ V provided by one of Arduino UNO's internal sources.

\begin{figure}
\begin{center}
\includegraphics[width=0.5\textwidth]{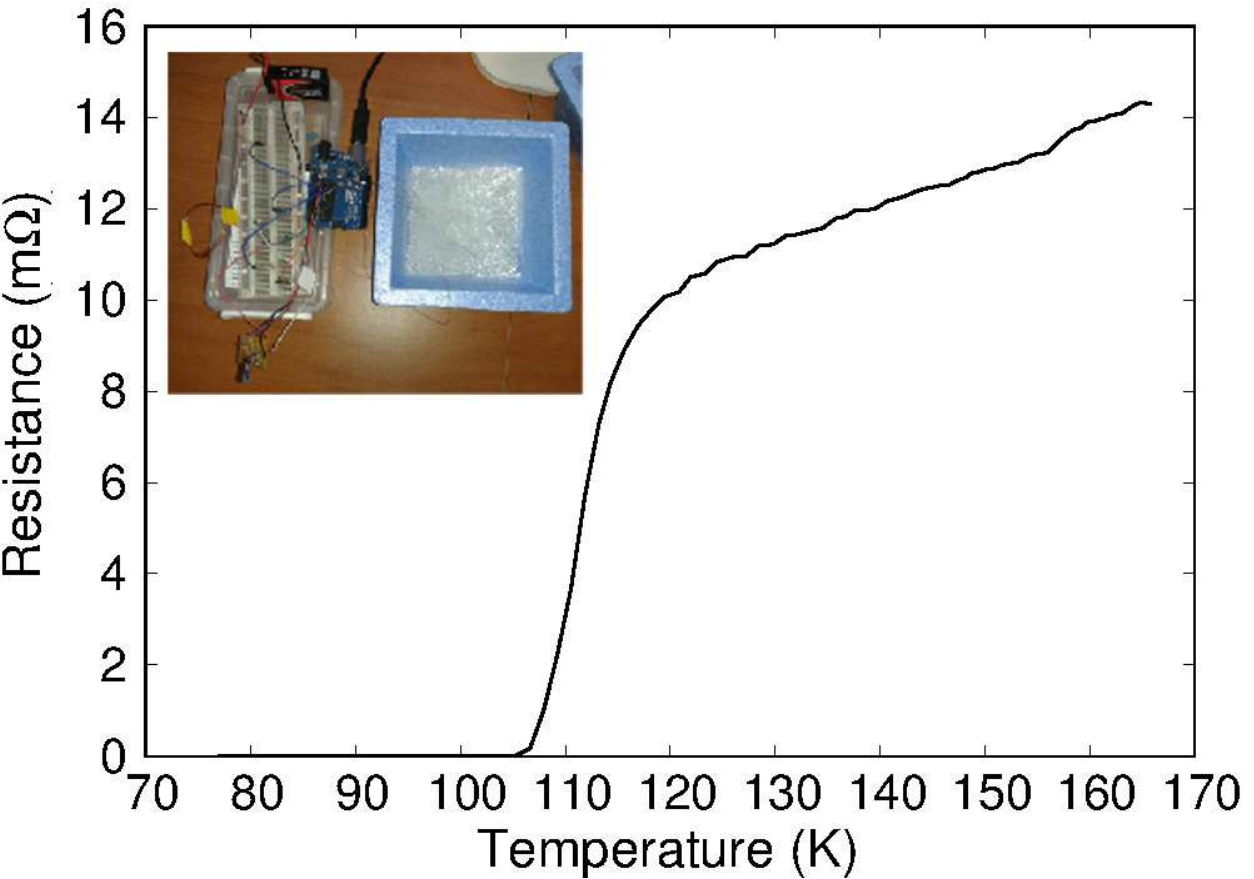}
\end{center}
\caption{Resistive transition of a superconductor measured with a voltage amplification. Inset: experimental setup. The blue polystyrene container is filled with glass beads. Both the superconductor and the Pt100 thermometer are immersed inside with liquid nitrogen.} \label{fig_4_point_ampli_supra}
\end{figure}

Both sample and thermometer have been attached to a printed circuit board and have been wrapped in cotton to ensure temperature homogeneity. The ensemble was placed in a polystyrene container filled with glass beads (inset of figure \ref{fig_4_point_ampli_supra}). Liquid nitrogen was then poured into the container and the temperature of the ensemble was let to increase back to room temperature while recording the data.

In this manner, students\cite{AL-AP} have measured the resistive transition  given in figure \ref{fig_4_point_ampli_supra}. The experimental data have been averaged by a convolution with a gaussian of half width 0.4 K to take into account the error on the temperature measurement. As can be seen, the resolution of the measurement is of the order of 0.1 m$\Omega$ for the superconductor's resistance. The latter is actually dominated by the thermal gradient that may exist between the sample and the thermometer if the operator does not carefully check that both are close to one another or if the container is forcefully warmed-up (with a hair dryer for instance). Nonetheless, the precision of the measurement is good ($<1\%$ relative uncertainty) and the measured mid-point $T_c$ is of 112 K $\pm$ 2K, very close to the value given by the specifications.

\section{Conclusion}
In conclusion, we have shown that standard temperature and resistance measurement methods could be adapted to microcontrollers. The performances that are then attainable are sufficient to probe with reasonable sensitivity a large range of physics phenomena such as thermoelectric effects, temperature-dependence of the resistivity, Hall effect, magnetocaloric effect,... We have illustrated this with the measurements of magnetocaloric effect in Gadolinium and of the resistive transition of a high critical temperature superconductor. We believe that the scope of inexpensive, transportable and easy-to-build experiments that are accessible through the use of single-board microcontrollers is continuously expanding and, in some cases, can now even replace standard characterization methods in research laboratories. They moreover provide a large range of opportunities to devise innovative teaching activities that involve enhanced students involvement.

\begin{acknowledgements}
We thank all the students who have participated in the Arduino-based labworks. We thankfully acknowledge Patrick Puzo for welcoming this project-based teaching within the Magist\`{e}re de Physique d'Orsay curriculum. This work has been supported by a ``P\'{e}dagogie Innovante'' grant from IDEX Paris-Saclay.
\end{acknowledgements}

\bibliography{Arduino_PIP}


\end{document}